\documentclass[10pt,twocolumn,superscriptaddress,floatfix,showpacs,nobalancelastpage,longbibliography]{revtex4-1}

\usepackage{array,longtable}
\newcolumntype{L}{>{\tiny $}p{0.33\columnwidth}<{$}}
\newcolumntype{M}{>{\scriptsize $}p{0.33\columnwidth}<{$}}
\newcolumntype{N}{>{\scriptsize $}p{0.43\columnwidth}<{$}}
\usepackage{amsmath}
\usepackage{amssymb}
\usepackage{amsfonts}
\usepackage{graphicx}
\usepackage{hyperref}

\allowdisplaybreaks[1]

\newif\ifhyper
\hypertrue
\ifhyper
\hypersetup{
   citecolor = {green},
   colorlinks = {true}, % false
   urlcolor = {blue} % magenta
}
\fi

\begin{document}

\title{Quantum critical phase with infinite projected entangled paired states}

\author{Didier Poilblanc and Matthieu Mambrini}
\affiliation{Laboratoire de Physique Th\'eorique, C.N.R.S. and Universit\'e de Toulouse, 31062 Toulouse, France}

\date{\today}

\begin{abstract}

A classification of SU(2)-invariant Projected Entangled Paired States (PEPS) on the square lattice, based on a unique site tensor,
has been recently introduced by Mambrini et al.~\cite{Mambrini2016}.
It is not clear whether such SU(2)-invariant PEPS can either i) exhibit long-range magnetic order (like in the N\'eel phase)
or ii) describe a genuine quantum critical point (QCP) or quantum critical phase (QCPh) separating two ordered phases.
Here, we identify a specific family of SU(2)-invariant PEPS of the classification which provides 
excellent variational energies for the $J_1-J_2$ frustrated Heisenberg model, especially at $J_2=0.5$, 
corresponding to the approximate location of the QCP or QCPh separating the N\'eel phase from a dimerized phase.
The PEPS are build from virtual states belonging to the $\frac{1}{2}^{\otimes N} \oplus 0$ SU(2)-representation, i.e. with $N$
``colors'' of virtual \hbox{spin-$\frac{1}{2}$}.
Using a full update infinite-PEPS approach directly in the thermodynamic limit, based on the Corner Transfer Matrix 
renormalization algorithm supplemented by a 
Conjugate Gradient optimization scheme, we provide 
evidence of i) the absence of magnetic order and of ii) diverging correlation lengths (i.e. showing no sign of saturation with increasing environment dimension) in both the singlet 
and triplet channels, when the number of colors $N\ge 3$. We argue that such a PEPS gives a qualitative description of the QCP or QCPh of the
$J_1-J_2$ model. 
  
 \end{abstract}
\maketitle

\section{Introduction}

\begin{figure}[htbp]
\begin{center}
\includegraphics[width=\columnwidth,angle=0]{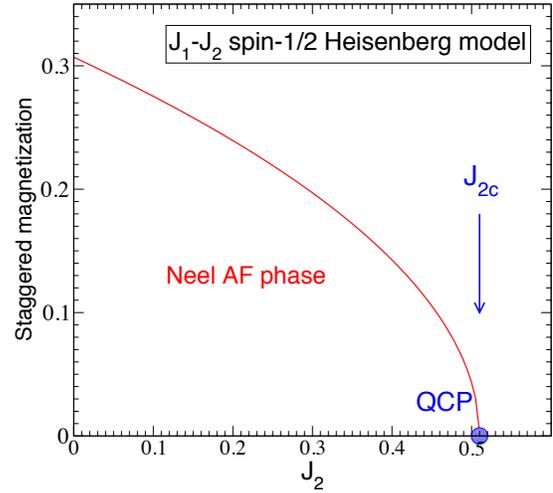}
\caption{[Color online] Schematical behavior of the staggered magnetization of the spin-$\frac{1}{2}$ $J_1-J_2$ Heisenberg model 
%Our analysis (lozenges) gives a critical value $J_{2c}\simeq 0.503(1)$, (crudely) assuming $m_{\rm stag}\sim m_0 \sqrt{J_{2c}-J_2}$
($J_1$ is set to 1). $m_{\rm stag}$ vanishes at the QCP. The exact location $J_{\rm 2c}$ of the QCP,
may be close to $0.5$. Recent DMRG studies~\cite{Wang2017} quote a narrow QCPh region around $J_2=0.5$.}
\label{FIG:PhaseDiag}
\end{center}
\end{figure}

% Generalities on quantum phase transitions and frustrated magnets. 
Low-dimensional quantum magnets offer a rich zoo of phases breaking a discrete (like point group or lattice) 
or a continuous (like spin rotation) symmetry.
Often, such phases are separated by Quantum Critical Points (QCP), as described within the usual Ginsburg-Landau (GL)
framework. 
Interestingly, it has been proposed that some QCP may {\it not} be described by the GL paradigm~\cite{Senthil2004,Shao2016}. 
A celebrated quantum spin model is the frustrated \hbox{spin-$\frac{1}{2}$} Heisenberg model on the two-dimensional (2D) square lattice 
involving competition between
nearest neighbor (NN) and next-nearest neighbor (NNN) antiferromagnetic (AF) couplings, $J_1$ and $J_2$ respectively. 
Setting $J_1=1$, $J_2$ controls the amount of frustration which is maximum (classically) at $J_2=0.5$. 
Large-scale Quantum Monte Carlo (QMC) simulations~\cite{Reger1988,Sandvik2010a,Sandvik2010b} has shown that the ground state (GS) of the unfrustrated ($J_2=0$) Heisenberg model exhibits
long range (LR) AF order.
In the thermodynamic limit, the (global) spin-rotational SU(2) symmetry is 
spontaneously broken  and the GS acquires a finite
local staggered magnetization. 
When $J_2$ is turned on, 
the order parameter is gradually suppressed and a quantum phase transition to a Quantum Disordered (QD) phase~\cite{Chandra1988,Dagotto1989a,Sachdev1991,Mila1991} -- such as a dimer~\cite{Sachdev1989,Dagotto1989b,Poilblanc1991,Schulz1996} or a plaquette~\cite{Zhitomirsky1996,Mambrini2006} Valence Bond Crystal (VBC) --
takes place (see Fig.~\ref{FIG:PhaseDiag}).  It was also argued that magnetic frustration could stabilize spin liquids (with no symmetry breaking),
such as the Resonating Valence Bond (RVB) states~\cite{Anderson1973}
showing algebraic (short range) VBC correlations on the square (Kagome) lattice~\cite{Albuquerque2010,Tang2011,Poilblanc2012,Schuch2012}. 

\begin{figure*}[htbp]
\begin{center}
\includegraphics[width=\textwidth,angle=0]{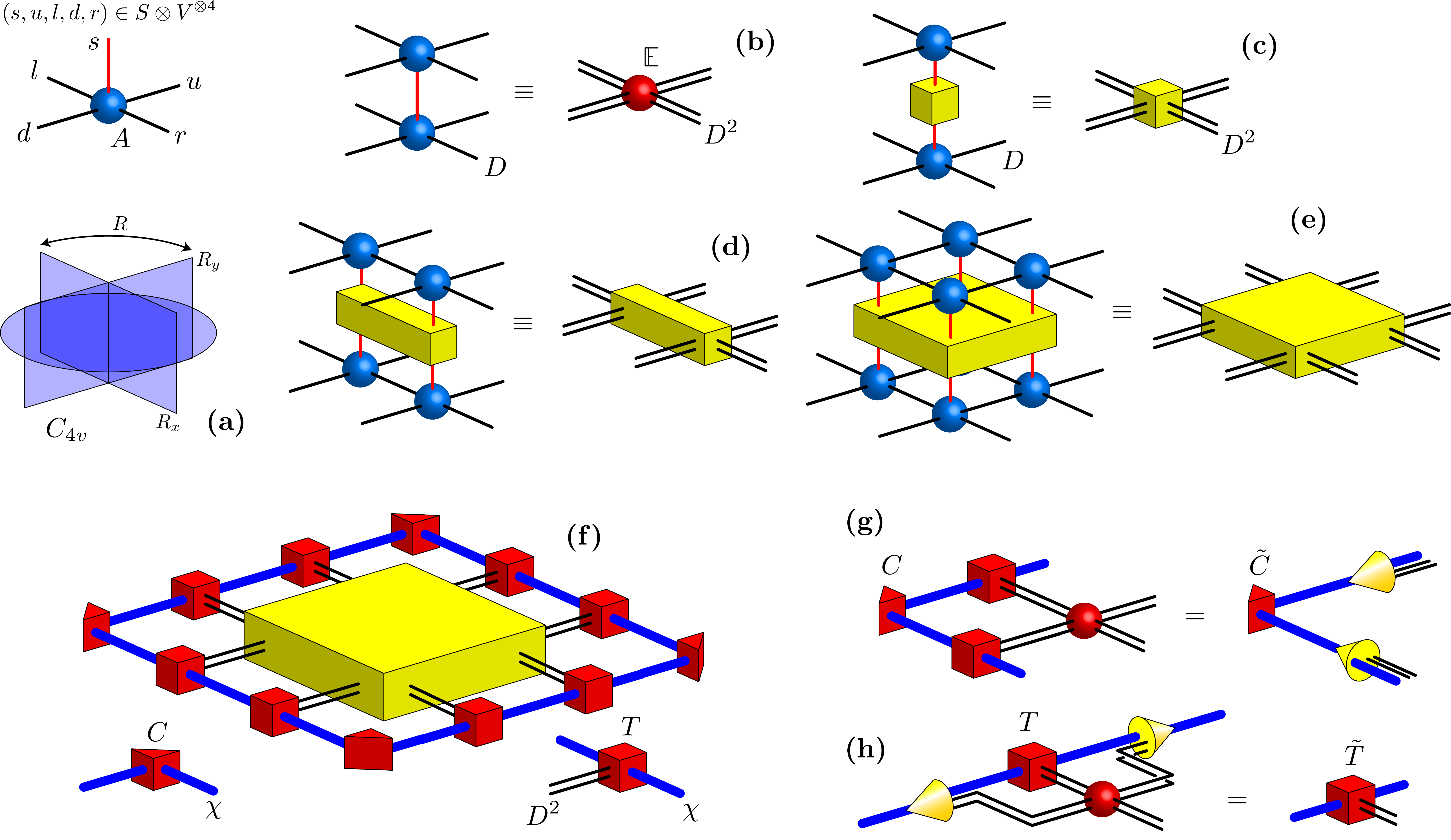}
\caption{[Color online] (a) Symmetric PEPS tensor $A$ with one 
physical index $s=\pm\frac{1}{2}$ and four virtual indices $u$, $l$, $d$ and $r$ 
(of dimension $D$).
$A$ is invariant under the generators of the $C_{4v}$ point group, i.e. the 90-degree rotation $R$, the reflection $R_x$ and the inversion $I=R_xR_y$.  
(b-e) The ``two layer'' (TL) tensors have bond dimension $D^2$ (double lines). 
One-site, two-site and four-site TL tensors obtained by inserting the identity $\mathbb I$, a one-site, a two-site
and a four-site operator, respectively.
(f) iPEPS CTM method~: a $2\times 2$ cluster is surrounded by a (self-consistent) environment build from a corner $\chi\times\chi$ transfer matrix $C$ and a side $\chi\times D^2\times \chi$ tensor $T$. In practice, we choose $\chi=k D^2$, $k\in\mathbb{N}$. Here the operator inserted on the 4-site is either ${\mathbb I}^{\otimes 4}$ (normalization) or the $J_1-J_2$ Hamiltonian. 
(g) Tensor renormalisation scheme~:  after one site is added, the new $\chi D^2\times \chi D^2$ CTM is diagonalized and only the largest (in modulus)
$\chi$ eigenvalues are kept to get the new CTM. 
(h) The unitaries approximated by isometries  (yellow pyramids) are used to compute the
new edge tensor.
}
\label{FIG:tensor}
\end{center}
\end{figure*}

%Generalities about tensor networks and PEPS. 
Recently, tremendous progress have been made in tensor network techniques~\cite{Cirac2009b,Cirac2012a,Orus2013,Schuch2013b,Orus2014}, aiming to go beyond Density Matrix Renormalization Group (DMRG) methods~\cite{White1992} in 2D.
More specifically, Projected Entangled Pair States (PEPS)~\cite{Perez-Garcia2007} are variational ans\"atze constructed from a few local tensors, located on $M$ non-equivalent sites, and characterized by 
(i) one bond carrying the physical degrees of freedom (of dimension $2$ for spin-$\frac{1}{2}$ systems) and (ii) $z$ ``virtual''
bonds ($z$ is the lattice coordination number, $z=4$ for the square lattice) of arbitrary dimension $D$ as shown in Fig.~\ref{FIG:tensor}(a).
Interestingly, any local (gauge) or global (physical) symmetry can be implemented in 
PEPS~\cite{Perez-Garcia2010,Schuch2010a,Singh2010a,Singh2012,Singh2013,Weichselbaum2012,Jiang2015,Haegeman2015,Mambrini2016}. 
Also, a simple bulk-edge (holographic) correspondence provides a remarkable tool to investigate the properties of edge states~\cite{Cirac2011,Yang2014}.
Many remarkable states of matter such as trivial paramagnets~\cite{Jian2016}, topological~\cite{Schuch2012,Poilblanc2012,Poilblanc2013a} or algebraic~\cite{Poilblanc2012} RVB spin liquids,  loop spin liquids~\cite{Li2014},
superfluids~\cite{Poilblanc2013b} or 
unconventional correlated superconductors~\cite{Poilblanc2014b} have simple representations in terms of PEPS. 
Numerical calculations with PEPS do not require to compute the wave function coefficients (which, conceptually, are given by contracting the tensor network over all virtual links) but, rather, make use of transfer matrices~\cite{Haegeman2016} based on  
 ``double-layer'' tensors (see Fig.~\ref{FIG:tensor}(b-e)). In the infinite-PEPS (iPEPS) method~\cite{Jordan2008}, one works directly 
 in the thermodynamic limit by approximating
the (infinite) space around a small $M$-site cluster by an effective ``environment" 
(see Fig.~\ref{FIG:tensor}(f)). One of the most accurate computation of the
environment is based on a Renormalization Group scheme involving 
Corner Transfer Matrices (CTMRG)~\cite{Nishino1996,Nishino2001,Orus2009,Orus2012}
as shown in Fig.~\ref{FIG:tensor}(g,h).
Unrestricted energy minimization over the $MdD^z$ tensor coefficients can be performed using
Time Evolution Block Decimation (TEBD)~\cite{Vidal2007b,Orus2008}
which has to be combined with a simple update~\cite{Vidal2003b,Jiang2008} or a full update~\cite{Phien2015} of the environment.  
A (finite) PEPS method using a $2\times 2$ cluster update 
supplemented by a finite size extrapolation has also been introduced~\cite{Wang2016}. 
Recently, a new optimization scheme using a Conjugate Gradient (CG) algorithm has been tested on the
non-frustrated~\cite{Corboz2016,Vanderstraeten2016} and frustrated~\cite{Liu2016} Heisenberg model, with iPEPS 
or finite PEPS, respectively.

% Motivations of the paper. SU(2) classification. 
The entanglement entropy (i.e. the quantity measuring the amount of entanglement in a bi-partitioned system) in a spontaneously-broken 
state exhibits anomalous additive logarithmic corrections~\cite{Kallin2011,Metlitski2011,Ju2012} to the area law (i.e. the linear scaling of the entropy with the length of the cut). 
When the staggered magnetization $m_{\rm stag}\rightarrow 0$, at the QCP, the violation of the area law is expected to be even more severe. 
This means that a good description of the QCP, or even of the N\'eel state, in terms of a PEPS (which strictly fulfills the area law for any finite $D$) is particularly challenging. 
A very simple ($D=3$) PEPS ansatz for the N\'eel state on the square lattice was first proposed in terms of a 
(one-parameter) spinon-doped RVB phase~\cite{Poilblanc2014a}. Also, finite size PEPS~\cite{Lubasch2014} or, more recently, state-of-the-art iPEPS calculations involving 
a Conjugate Gradient (CG) minimization algorithm~\cite{Corboz2016,Vanderstraeten2016} came up with very 
accurate energy for the
N\'eel GS of the 2D Heisenberg model. However, the phase diagram of the $J_1-J_2$ model is
still heavily debated.
No agreement has been reached between several numerical approaches, neither on the nature of  the QD region -- with proposals of VBC~\cite{Schulz1996,Mambrini2006,Capriotti2000,Gong2014}, (topological) gapped~\cite{Jiang2012}
or gapless~\cite{Capriotti2001,Wang2013,Hu2013,Gong2014,Morita2015} spin liquids -- nor on the location $J_2=J_{2c}$ of the phase transition.
While early Exact Diagonalisations (ED) extrapolations~\cite{Schulz1996} were bracketing  $J_{2c}\in [0.34,0.6]$,
DMRG studies~\cite{Jiang2012,Gong2014} suggested $J_{2c}\simeq 0.41 - 0.44$, while Variational Monte Carlo (VMC)
studies~\cite{Hu2013,Morita2015} give $J_{\rm 2c}\simeq 0.48 - 0.5$ and finite-size (cluster update) PEPS 
computations~\cite{Wang2016} $J_{2c}\simeq 0.572(5)$. Recently, Wang and Sandvik~\cite{Wang2017} argued for a quantum critical phase (QCPh) centered around 0.5. 
In all these approaches (except ED), the spin rotational SU(2) symmetry is {\it explicitly} broken in the 
N\'eel phase.
However, there is no obstruction principle to construct accurate SU(2)-symmetric wave functions exhibiting 
long range AF order~\cite{Kaneko2016}. Since such states may  be characterized by a large entanglement, it is unclear
whether it can be realized with low-$D$ symmetric PEPS. 
Also, whether SU(2)-symmetric PEPS
have the potential to describe zero-temperature QCP or QCPh --  in the same 
way as one-dimensional (1D) Matrix Product States (MPS) can describe 
critical 1D systems~\cite{Tagliacozzo2008,Pollmann2009,Pirvu2012} -- is still unclear~\cite{Liu2010}.
Though, it is known that non-trivial criticality can be captured by PEPS, even at finite $D$~\cite{Nishino2001,Verstraete2006a}.

Motivated by the above conceptual and practical issues, we have re-visited the $J_1-J_2$ model 
using some new PEPS developments,
based on a general scheme to construct SU(2)-symmetric PEPS using computer-assisted algebra~\cite{Mambrini2016}.
This enables us to introduce key features in the full-update iPEPS scheme~:
(i) Full translational and rotational invariance is enforced by using a unique SU(2)-invariant tensor on every lattice site; 
(ii) Full optimization of the (few) tensor coefficients is accomplished via a CG method;
(iii) Careful scaling with environment dimension $\chi$ is performed in order to address the $\chi\rightarrow\infty$ limit.
Using this procedure, we have identified a specific (low-dimensional) family of SU(2)-symmetric PEPS
which provides 
excellent variational energies for the $J_1-J_2$ frustrated Heisenberg model, especially at $J_2=0.5$, 
i.e. close to the (unknown) QCP or QCPh of this model.
We show evidence that these (optimized) PEPS do not exhibit long range AF order.
We also find that, above bond dimension $D=7$, the PEPS (optimized for $J_2=0.5$) exhibits diverging spin-spin and dimer-dimer correlation lengths, i.e. showing no sign of saturation up to large environment dimension.
In addition, a small spurious $m_{\rm stag}$ is found to vanish in the limit of infinite environment dimension. 
Hence, we propose that this state offers a realization of the QCP or QCPh.

\section{Symmetric PEPS ans\"atze} 

We wish here to consider transitionally invariant
fully symmetric PEPS in order to (i) reduce the number of independent variational parameters and (ii) provide a good description of the critical point (or phase) where both SU(2) and lattice symmetries are preserved.
For this purpose, we shall use the elegant classification of SU(2)-invariant PEPS tensors on the square lattice~\cite{Mambrini2016} according to (i) their virtual degrees of freedom and (ii) how they transform w.r.t the (lattice) point group symmetries (see Fig.~\ref{FIG:tensor}(a)).
For simplicity, we shall a priori restrict ourselves to tensors fully invariant under all operations of the $C_{4v}$ point group (i.e. belonging to the so-called $A_1$ IRREP). The tensors are further classified
according to their virtual space $V$ given by a direct sum of SU(2) IRREPs or ``spins'', i.e.
$V=\bigoplus_\alpha s_\alpha$. 
We restrict hereafter to bond dimension $D\leq 7$. Among all the possible cases listed in Table \ref{TABLE:classes} we focus on the most interesting ones carrying low virtual spins defined by
$V=\frac{1}{2}\oplus 0$ ($D=3$), $V=\frac{1}{2}\oplus 0\oplus 0$ ($D=4$), $V=1\oplus\frac{1}{2}$ ($D=5$), $V=\frac{1}{2}\oplus\frac{1}{2}\oplus 0$ ($D=5$), 
$V=\frac{1}{2}\oplus\frac{1}{2}\oplus\frac{1}{2}\oplus 0$ ($D=7$), spanned by 
a small number $\cal D$ of independent tensors, ${\cal D}=2, 8, 4,10,30$ respectively, given in the Supplementary Materials of
Ref.~\onlinecite{Mambrini2016} (except for $D=7$ given in the Supplementary Materials of this paper~\cite{SM2017}). 
Note that a $\pi$-rotation of the spin basis is assumed on the sites of one of the two sublattices of the square lattice. 
In this basis, a genuine ${\bf q}={\bf q}_{\rm AF}\equiv (\pi,\pi)$ (spontaneous) magnetic order translates into a {\it uniform} ${\bf q}=0$ (spontaneous) magnetization. Subsequently, the generator of $SU(2)$ become invariant only up to translations that map the 
sublattices to themselves (i.e. shifts over two sites).

%%%%% Added Resub %%%%%%%%%
\begin{table}[]
\centering
\begin{tabular}{lp{1mm}lp{1mm}lp{1mm}llp{1mm}l}
\hline\hline
\rule{0pt}{0ex} $D$ & \multicolumn{2}{c}{3}                       & \multicolumn{2}{c}{4}                             & \multicolumn{2}{c}{5}                                        & 6                                                & \multicolumn{2}{c}{7}                                                          \\ \hline
\rule{0pt}{2.5ex} $V$ & {\textcolor{green}{\checkmark}} &$\frac{1}{2} \oplus 0$ & {\textcolor{green}{\checkmark}} &$\frac{1}{2}\oplus 0 \oplus 0$ &  &$\frac{1}{2}\oplus 0 \oplus 0\oplus 0$   & $1\oplus \frac{1}{2}\oplus 0$  &  {\textcolor{green}{\checkmark}} &$\frac{1}{2}\oplus \frac{1}{2}\oplus \frac{1}{2} \oplus 0$    \\
\rule{0pt}{2.5ex}    &   &     &                &                                & {\textcolor{green}{\checkmark}} &$\frac{1}{2}\oplus \frac{1}{2} \oplus 0$ &  &  &$1\oplus \frac{1}{2}\oplus 0\oplus 0$  \\
\rule{0pt}{2.5ex}    &   &     &                &                                & {\textcolor{green}{\checkmark}} &$1\oplus \frac{1}{2}$                    &                     & &$\frac{3}{2}\oplus \frac{1}{2}\oplus0$ \\
\rule{0pt}{2.5ex}    &   &     &                &                                & &$\frac{3}{2}\oplus  0$                   &                    &        &$\frac{3}{2}\oplus 1$               \\
\rule{0pt}{2.5ex}    &   &     &                &                                & &                                         &                                                  &     &$2\oplus\frac{1}{2}$                \\
\rule{0pt}{2.5ex}    &   &     &                &                                & &                                         &                                                  &   &$\frac{5}{2}\oplus 0$                   \rule[-1.5ex]{0pt}{0pt}
 \\ \hline\hline
\end{tabular}
\caption{[Color online] List of all virtual spaces $V$ of bond dimension $D\le 7$ for which $V^{\otimes 4}$ can be projected onto a physical spin 1/2. The ones considered here are indicated by (green) marks. Classes with higher spins give poorer variational energies than the lower spin ones 
of same total bond dimension $D$. }
\label{TABLE:classes}
\end{table}

{\it The iPEPS method combined with full tensor optimization} -- 
We shall now focus on the $J_1-J_2$ spin-$\frac{1}{2}$ Heisenberg model with NN and NNN
 antiferromagnetic coupling $J_1$ and $J_2$, respectively,
which we have studied at $J_2=0$ in the absence of frustration and, for strong frustration,
at $J_2=0.5$ and $J_2=0.55$. Our first goal is to optimize the variational energy within each 
$\cal D$-dimensional class of SU(2)-invariant PEPS i.e finding the optimum linear superposition of the
$\cal D$ independent tensors of each class. Since the number of variational parameters remains small
(maximum of ${\cal D}=30$ for $D=7$) we have used a "brute force" CG optimization as e.g. given in Numerical Recipes~\cite{Numerical2007}. However, this requires an efficient iPEPS computation of the variational energy  for any set of variational parameters to ``feed'' the CG routine. This is performed constructing a self-consistent environment around an active $2\times 2$ cluster 
(see Fig.~\ref{FIG:tensor}(b)) using an iterative CTMRG algorithm~\cite{Nishino1996,Orus2009,Orus2012} 
optimized for spatially symmetric tensors. Indeed, we have introduced simple modifications:
(i) we use a {\it unique} CTM $C$ tensor (side tensor $T$) which is the same for all
corners (edges) and (ii) the basic Singular Value Decomposition (SVD) in each CTMRG step 
to construct the environment is replaced by a (more stable) ED, the CTM being 
here a symmetric matrix. The largest environment dimension we could handle was $\chi=400$ and 
$\chi=294$ for $D=5$ and $D=7$, respectively, for which up to 350 or 400 iterations became necessary to converge the environment. 
Note that the initial $C$ ($T$) tensor is obtained from the $\mathbb{E}$ tensor of Fig.~\ref{FIG:tensor}(b) by summing over all external
$l$ and $u$ ($u$) indices. 

\begin{figure}[htbp]
\begin{center}
\includegraphics[width=\columnwidth,angle=0]{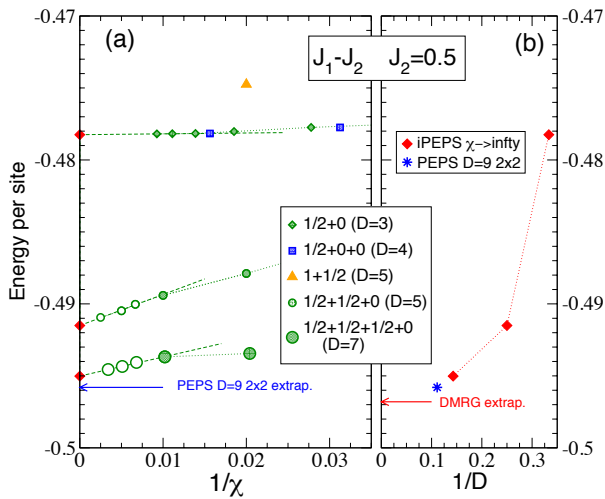}
\caption{[Color online] (a) iPEPS variational energies of the $J_1-J_2$ model
at $J_2=0.5$, versus the
inverse of the environment dimension $\chi$. Full (open) symbols correspond to fully optimized (fixed) tensor
ans\"atze (see text). $\chi\rightarrow\infty$ linear extrapolations are performed using only the last data points. 
(b) Behavior of the $\chi\rightarrow\infty$ extrapolated energies vs the inverse of the bond dimension $D$.
$D=9$ PEPS~\cite{Wang2016} and DMRG~\cite{Gong2014} {\it extrapolated} energies are shown for comparison
(see also Table~\protect\ref{TABLE:comparisons}).}
\label{FIG:energy}
\end{center}
\end{figure}

{\it Energetics} --
Variational energies (per site) in each class of tensors are shown in Fig.~\ref{FIG:energy}(a) for $J_2=0.5$,
as a function of the inverse of the environment dimension $\chi$. 
A rapid comparison between the different classes (for intermediate $\chi$) reveals that, for identical bond dimension $D$, 
the classes $V=\frac{1}{2}^{\otimes N}\oplus 0$ with $N=1$, $2$ and $3$ (of bond dimensions $D=3$, $5$ and $7$, respectively) 
give the best results. Hence, hereafter we shall focus on this PEPS family defined in terms of $N$ ``colors'' of spin-$\frac{1}{2}$. Note that the case $N=1$ was studied previously in Ref.~\cite{Wang2013}.
Tensors are fully optimized up to a maximum bond dimension $\chi_{\rm opt}$,
e.g. for $D=5$, $\chi_{\rm opt}=4D^2=100$ and, for $D=7$, $\chi_{\rm opt}=2D^2=98$. Then, using environment dimensions $\chi>\chi_{\rm opt}$ together with the fixed optimized tensor obtained at $\chi=\chi_{\rm opt}$, one gets true {\it upper bounds} of the variational energy. 
In contrast, for $D=3$, $\chi_{\rm opt}=12D^2=108$ already gives the absolute best tensor with enough accuracy. 
Generically, we found that the energy always decreases with increasing $\chi$ and, at large enough $\chi$, linear fits can be performed in $1/\chi$ to provide  $\chi\rightarrow\infty$ extrapolations, also upper bounds of the ($D$-dependent) 
variational energies. Note that our $D=7$ extrapolation $-0.49502$ lies within
only $0.2\%$ of the extrapolated value $-0.4958$ obtained using cluster update finite size $D=9$ PEPS~\cite{Wang2016}.
We have plotted our ($\chi\rightarrow\infty$) results as a function of $1/D$ in Fig.~\ref{FIG:energy}(b) showing 
perfect consistency with the above-mentioned $D=9$ result together with the DMRG extrapolation $-0.4968$ of Ref.~\onlinecite{Gong2014}.
This agreement is remarkable considering the fact that we use only a unique tensor parametrized by a small number
of coefficients. 
Good variational energies have also been found for the simple NN Heisenberg model ($J_2=0$) as 
well as for larger frustration $J_2=0.55$ as shown in Appendix~\ref{app1}.
Our results are summarized in Table~\ref{TABLE:comparisons} and compared to the best estimates, from Quantum Monte Carlo
at $J_2=0$~\cite{Sandvik2010a,Sandvik2010b} and from DMRG~\cite{Gong2014}, VMC~\cite{Hu2013} 
and finite-size PEPS~\cite{Wang2016} at $J_2=0.5$ and $J_2=0.55$.
We note however that our variational energies for $J_2=0$ and $J_2=0.55$ are slightly less accurate as for $J_2=0.5$.
In fact, we believe $J_{2c}$ is close to 0.5 and we argue below that our (optimized) PEPS is capable of picking 
up the critical nature of the QCP or QCPh.
For $J_2=0.55$ translation symmetry breaking is likely to occur spontaneously, which is not captured by our homogeneous ansatz. The ansatz does not either sustain magnetic LR order, that may explain its lower accuracy at $J_2=0$.

\begin{table}[htbp]
\begin{center}
\resizebox{0.9\columnwidth}{!}{%
%\resizebox{\columnwidth}{!}{%
  \begin{tabular}{@{} cccc @{}}
\hline
    \hline 
     J & 0 & 0.5 & 0.55  \\ 
  \hline 
 QMC &  $-0.66944$ &  &  \\
DMRG &  & $-0.4968$ & $-0.4863$ \\
VMC   & & $-0.4970(5)$ & $-0.4870(5)$  \\
$D=9$ PEPS &   & $-0.4958(3)$ & $-0.4857(2)$ \\
 $D=7$ iPEPS &  $-0.6677$ & $-0.4950$ & -0.4830  \\
    \hline 
      \end{tabular}
     }
\caption{Comparison between our $D=7$ iPEPS results ($\chi\rightarrow \infty$ extrapolations) and the best estimates
in the literature, for $J_2=0$, $J_2=0.5$ and $J_2=0.55$~: $J_2=0$ results are obtained by QMC~\cite{Sandvik2010a,Sandvik2010b}.
At finite $J_2$, we quote energies
obtained by extrapolations to the thermodynamic limit using DMRG~\cite{Gong2014}, VMC~\cite{Hu2013} and 
{\it finite-size} $D=9$ PEPS~\cite{Wang2016}. Note that the $D=7$ iPEPS energies are only upper bounds of the true 
variational energies (see text).
}
\label{TABLE:comparisons}
\end{center}
\end{table}

\section{Correlation functions}

Once the PEPS $|\Psi_0\big>=|\Psi (D,\chi_{\rm opt}) \big>$ has been optimized using the largest 
possible environment dimension $\chi=\chi_{\rm opt}(D)$, various correlation functions can be computed 
(e.g. along the ${\bf e}_x$ horizontal direction), like (i) the spin-spin correlations,
\begin{equation}
C_{\rm s}(d)=\big< {\bf S_i}\cdot{\bf S}_{{\bf i}+d {\bf e}_x}\big>_0\, ,
\label{EQ:ss}
\end{equation}
(ii) the (connected) {\it longitudinal} dimer-dimer correlations,
\begin{equation}
C_{\rm d}^{(\rm L)}(d)=\big< D_{\bf i}^x D_{{\bf i}+d {\bf e}_x}^x\big>_0
- \big< D_{\bf i}^x \big>_0  \big<D_{{\bf i}+d {\bf e}_x}^x\big>_0 \, ,
\label{EQ:ldd}
\end{equation}
and (iii) the (connected) {\it transverse} dimer-dimer correlations,
\begin{equation}
C_{\rm d}^{(\rm T)}(d)=\big< D_{\bf i}^y D_{{\bf i}+d {\bf e}_x}^y\big>_0
- \big< D_{\bf i}^y \big>_0  \big<D_{{\bf i}+d {\bf e}_x}^y\big>_0 \, ,
\label{EQ:tdd}
\end{equation}
where dimer operators $D_{\bf i}^x={\bf S_i}\cdot{\bf S}_{{\bf i}+{\bf e}_x}$ and 
$D_{\bf i}^y={\bf S_i}\cdot{\bf S}_{{\bf i}+{\bf e}_y}$
are oriented either along the ${\bf e}_x$ (horizontal) or ${\bf e}_y$
(vertical) directions, respectively, and the expectation values are taken in the optimized
$|\Psi_0\big>$ PEPS. 

\begin{figure*}[htbp]
\begin{center}
\includegraphics[width=\textwidth,angle=0]{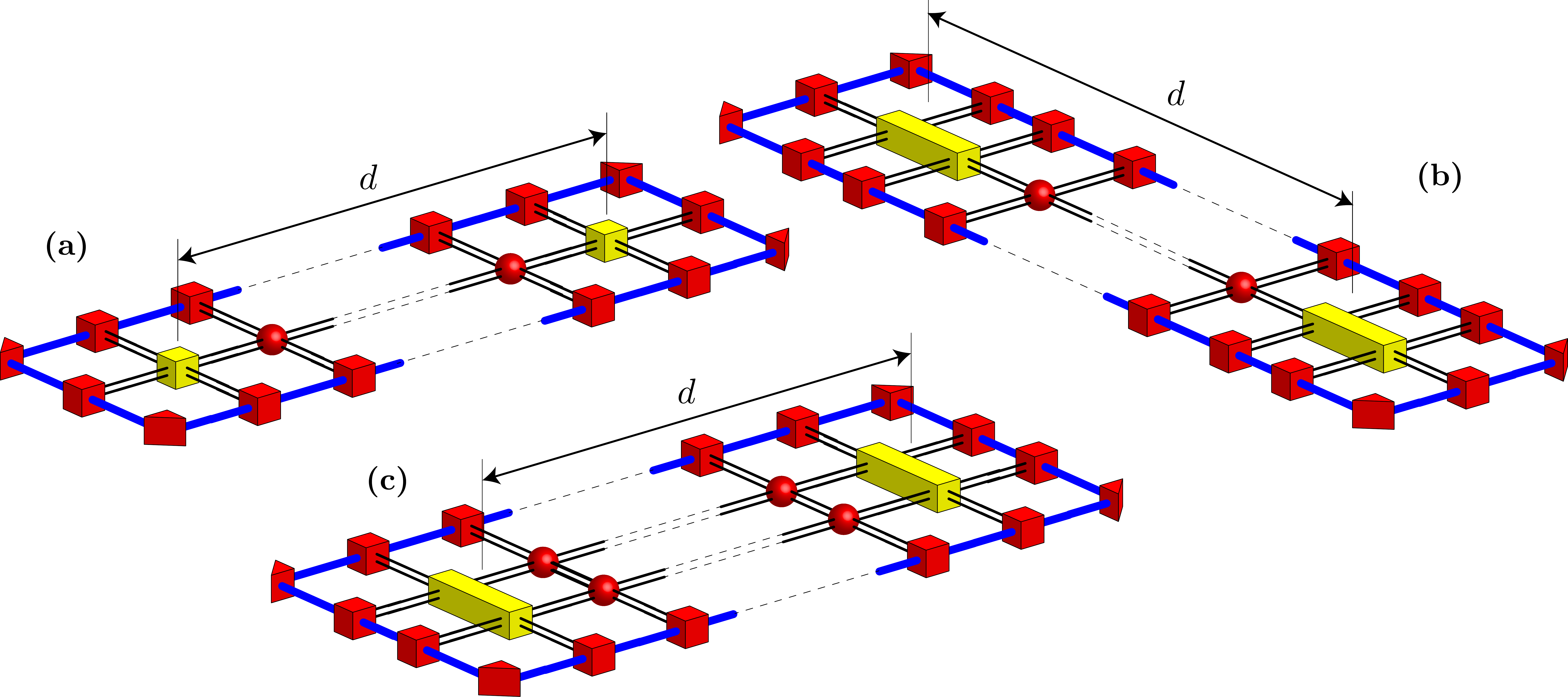}
\caption{[Color online] 
One dimensional strips used to compute the spin-spin (a), the longitudinal dimer-dimer (b) and the transverse dimer-dimer (c) correlation functions. A transfer matrix is applied recursively $d-1$ times (a,c) or $d-2$ times (b) in the direction of the strip. }
\label{FIG:correlators}
\end{center}
\end{figure*}

\begin{figure}[htbp]
\begin{center}
\includegraphics[width=\columnwidth,angle=0]{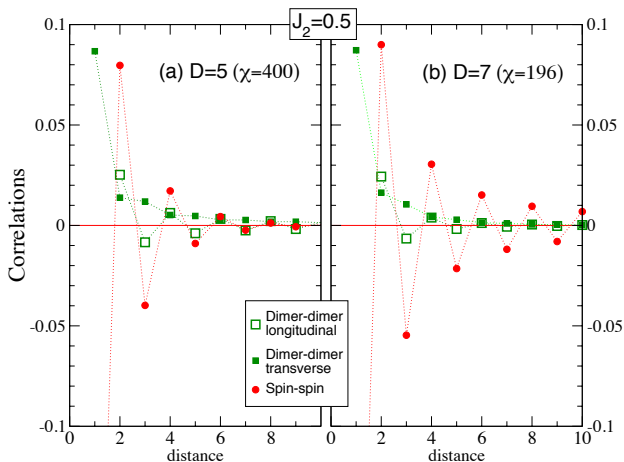}
\caption{[Color online] Short-distance correlation functions at $J_2=0.5$ 
for $V=\frac{1}{2}\oplus\frac{1}{2}\oplus 0$ (a) 
and $V=\frac{1}{2}\oplus\frac{1}{2}\oplus\frac{1}{2}\oplus 0$ (b).
Large environment dimensions $\chi$ are used ensuring full convergence of the correlations at short distance
($r<10$). }
\label{FIG:short}
\end{center}
\end{figure}

The calculations of correlators are accomplished using the set-up shown in Fig.~\ref{FIG:correlators}(a-c). 
Appropriate transfer matrices are used so that one can construct arbitrarily long strips.
Here the site tensor is fixed to its optimized output using $\chi=\chi_{\rm opt}(D)$ (hereafter we use $\chi_{\rm opt}=49$ for $D=7$) while  the environment dimension $\chi>\chi_{\rm opt}(D)$ can be then further increased to reach convergence,
which is easily achieved for short distance $r$.  A comparison between the results obtained with the two ans\"atze 
$V=\frac{1}{2}\oplus\frac{1}{2}\oplus 0$ (a) 
and $V=\frac{1}{2}\oplus\frac{1}{2}\oplus\frac{1}{2}\oplus 0$ (b) is shown in Fig.~\ref{FIG:short} for $J_2=0.5$. 
Although a fast decay of the dimer-dimer 
correlations is seen in both cases, the behavior of the (staggered) spin-spin correlations is qualitatively 
different~: for $D=7$ $|C_{\rm s}(r)|$ seems to approach a finite value  while, for $D=5$ (or $D=3$ as well),
it steadily decays to zero. This signals the emergence, for $D\ge7$, 
of a finite staggered magnetization as defined by
$
m_{\rm stag}(\chi)=\sqrt{\lim_{r\rightarrow\infty}|C_{\rm s}(r)|}\, .
$
We note however that, strictly speaking, for finite $\chi$ the above limit should vanish since the correlations are cut-off 
above some correlation length $\xi_{\rm s}(\chi)$ (see below). In other words, the strip of Fig.~\ref{FIG:correlators}(a) is,
crudely speaking, similar to a quasi-1D physical strip (ladder) of effective width $L_{\rm eff}(\chi)$~\cite{Nishino1996b},
which can not 
sustain long-range magnetic order from Mermin-Wagner theorem (MWT)~\cite{Mermin1966}.
However, MWT may not, strictly speaking, apply to a transfer operator as for a true Hamiltonian.
In addition, for $D=7$ the SU(2) symmetry is spontaneously broken~: 
small deviations from a perfectly SU(2)-symmetric environment act as a small symmetry-breaking (AF) "field" 
and  the local spin operator acquires a finite value $\big< {\bf S_i}\big>_0=\cos{({\bf q}_{\rm AF}\cdot{\bf i})} 
\,{\bf m}_{\rm stag}$ oscillating at the antiferromagnetic wave vector ${\bf q}_{\rm AF}$.
As shown in Appendix \ref{app2}, 
$m_{\rm stag}(\chi)$ vanishes in the $\chi\rightarrow\infty$ limit, physically corresponding to the
limit of an infinitely wide strip $L_{\rm eff}\rightarrow\infty$.
This implies 
that the infinite 2D system recovers the full SU(2) spin symmetry encoded in the tensor ansatz. 
We have seen similar behaviors also for $J_2=0$ and $J_2=0.55$ as well (see Appendix~\ref{app2}).
Interestingly, the scaling of $m_{\rm stag}$ to zero may depend slightly of the initial CTM of the CTMRG 
procedure to converge the environment. In contrast, for $D=3$ and $D=5$ 
the system remains spin isotropic even for finite $\chi$, the spin correlators 
$\big< S_{\bf i}^\alpha S_{\bf j}^\alpha\big>_0$ being independent on $\alpha=x,y,z$, as checked explicitly.
This signals a qualitative change of behavior when $N\ge 3$ which we identify in the next section.

\begin{figure}[htbp]
\begin{center}
\includegraphics[width=\columnwidth,angle=0]{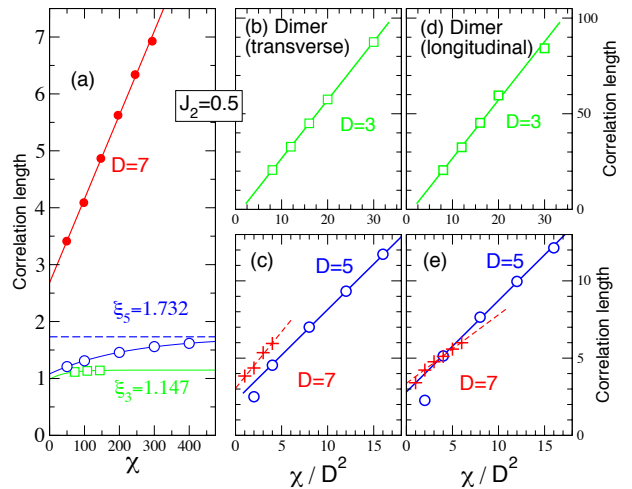}
\caption{[Color online] ) Scaling of the various correlation lengths at $J_2=0.5$ vs (a) environment dimension $\chi$
or (b)-(e) $\chi/D^2$, 
for $V=\frac{1}{2}\oplus 0$ (open squares), $V=\frac{1}{2}\oplus\frac{1}{2}\oplus 0$ (open circles) 
and $V=\frac{1}{2}\oplus\frac{1}{2}\oplus\frac{1}{2}\oplus 0$ (large dots and crosses).
(a) Spin-spin correlations; (b,c) Transverse dimer-dimer correlations; (d,e) Longitudinal dimer-dimer correlations.
\label{FIG:length}}
\end{center}
\end{figure}

{\it Diverging correlation lengths} --
The results described above give some hints that, when $D=7$, the spin-spin correlations become
algebraic at long distance. 
However, for finite bond dimension $\chi$, the strips of Fig.~\ref{FIG:correlators}(a-c) can be seen as
effective 1D systems. Then, finite correlation lengths $\xi_D(\chi)$ naturally emerge as the inverse of the gaps  of {\it finite-dimensional} $D_{\rm eff}^2\times D_{\rm eff}^2$ transfer matrices, where 
$D_{\rm eff}=D\chi$ (Fig.~\ref{FIG:correlators}(a,b)) or $D_{\rm eff}=D^2\chi$ (Fig.~\ref{FIG:correlators}(c)) 
are the effective dimensions of the associated 1D MPS. Using empirical findings for the correlation
length $\xi_{\rm 1D}$ in critical 1D systems~\cite{Tagliacozzo2008,Pollmann2009,Pirvu2012}, 
 $\xi_{\rm 1D}(D)\sim D^\kappa$, one then expects that $\xi_D(\chi)\sim (D_{\rm eff})^\kappa$, $\kappa>0$, which should diverge with $\chi$ as a power law for critical PEPS.
Hence,
criticality (if any) is restored only in the $\chi\rightarrow\infty$ limit and finite-$\chi$ scaling is necessary to
obtain informations on the QCP or QCPh.  
Note that, when spin rotational symmetry is (artificially) broken at finite $\chi$, it is important to consider the {\it connected}
spin-spin correlator ${\tilde C}_{\rm s}(d)=C_{\rm s}(d)-(m_{\rm stag})^2$.
From straightforward fits of the
 long-distance correlations at $J_2=0.5$ (see Appendix~\ref{app3}) we have extracted the correlation lengths $\xi_D(\chi)$ associated to the
 ${\tilde C}_{\rm s}$, $C_{\rm d}^{(\rm T)}$ and $C_{\rm d}^{(\rm L)}$ correlation functions and results are shown in Fig.~\ref{FIG:length}.
For $D=3$ or $D=5$ we find a clear saturation of the spin-spin correlation lengths to small values 
while the
dimer-dimer correlations lengths diverge {\it linearly} with $\chi$. Such a behavior is typical of bi-partite
dimer models~\cite{Sandvik2006} or of the NN RVB state on the square lattice~\cite{Albuquerque2010,Poilblanc2012} due to $U(1)$-gauge symmetry. 
In fact, the $D=3$ PEPS
can be viewed as an extended-range RVB state~\cite{Wang2013} and the $D=5$ PEPS as 
an extended-range {\it two-color} RVB state.
Plotting the dimer correlation lengths in Fig.~\ref{FIG:length}(c,e) as a function of $\chi/D^2$ clearly
reveals the similarities between $D=5$ and $D=7$. 
However, in the case of the spin correlations, a sudden qualitative change occurs at $D=7$ for which 
we find that the spin-spin correlation length no longer saturates but increases linearly with $\chi$, as the dimer correlation 
lengths do (see Appendices~\ref{app3} and \ref{app4} for details). No sign of saturation of the correlation lengths is observed up to the largest available environment dimensions.
This suggests that the (optimized) $D=7$ PEPS is critical in the limit $\chi\rightarrow\infty$ or, at least, can very well describe a critical state. 

\begin{figure}[htbp]
\begin{center}
\includegraphics[width=\columnwidth,angle=0]{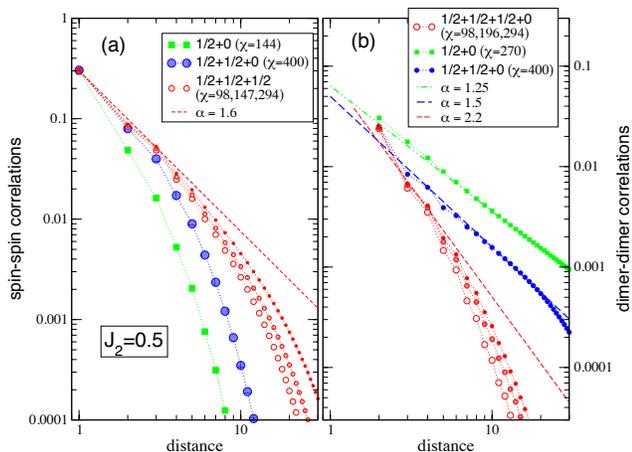}
\caption{[Color online] (a) Log-log plot of spin-spin (a) and longitudinal dimer-dimer (b) correlations 
versus distance. Straight (dashed) lines correspond to power-law \hbox{decays $\sim d^{-\alpha}$}. }
\label{FIG:loglog}
\end{center}
\end{figure}

{\it Power-law exponents} --
Whenever the correlation length $\xi_D(\chi)$ diverges (or becomes very large), 
one expect to see power-law behaviors in the correlation functions,
\begin{eqnarray}
C_{\rm s}(d)&\sim& d^{-(1+\eta_{\rm s})}\, ,\\
C_{\rm d}(d)&\sim& d^{-(1+\eta_{\rm d})}\, ,
\end{eqnarray}
in the range of distance
\hbox{$1< d < \xi_D$}, where $\eta_{\rm s}$ and $\eta_{\rm d}$ defined e.g. in Ref.~\cite{Lou2009}
are the anomalous dimensions.
Note however that this scaling regime can be observed only when $\xi_D(\chi)$ has reached a sufficiently large value.
 To obtain estimates of the exponents $1+\eta_{\rm s}$ and $1+\eta_{\rm d}$ we have plotted spin-spin and (longitudinal) dimer-dimer correlations at $J_2=0.5$ in Fig.~\ref{FIG:loglog}(a,b) using log-log scales. 
For $D=3$ ($D=5$) the dimer correlation length is very large (is large) for the largest $\chi$ we can reach and, from fits of the data
in the range $1<d<100$ ($1<d<20$), one can easily 
extract the exponent $1+\eta_{\rm d}\simeq 1.25$ ($1+\eta_{\rm d}\simeq 1.5$). 
For $D=7$, it is difficult to extract accurate exponents since cross-overs to exponential decays occur rapidly around $d\sim\xi_7\simeq 6$,
for both the spin-spin and dimer-dimer correlations. 
However, the systematic trend of the data with $\chi$ in Fig.~\ref{FIG:loglog}(a,b) suggests $\eta_{\rm s}\sim 0.6$ 
and $\eta_{\rm d}\sim 1.2$.  

{\it Discussion and outlook} --
Above, we have found solid evidence that the $N=3$ ($D=7$) SU(2)-invariant state exhibits slowly (possibly power law) decaying spin-spin and 
dimer-dimer correlation functions, suggesting a critical behavior, or at least very large correlation lengths.
We now argue that the family of SU(2)-symmetric tensors characterized by the virtual 
space $V=\frac{1}{2}^{\otimes N}\oplus 0$ with $N\ge 3$ ``colors'' can describe faithfully the QCP or QCPh of the
spin-$\frac{1}{2}$ $J_1-J_2$ Heisenberg model. 

First, we observed that spin-spin correlations decay less and less rapidly for increasing 
$N$ (i.e. $D$) so we expect such correlations to become longer and longer range for increasing $N$. Since the anomalous 
dimension $\eta_{\rm s}$ (defined from the correlation at intermediate distances) generically decreases
with increasing $D$, one can put an upper bound to its infinite-$D$ limit, namely $\eta_{\rm s}< 0.6$.

Secondly, it is remarkable that dimer-dimer correlations (and correlation lengths) become very similar for $N=2$ and $N=3$, if compared at the same value of the ratio $\chi/D^2$. In fact, we may speculate that, for $N\ge 3$, all correlation lengths diverge as
\begin{equation}
\xi_D(\chi)\simeq f_D\, \chi/D^2\, ,
\label{EQ:length}
\end{equation}
where the prefactor $f_D$ depends weakly on $D$, the main effect of increasing the bond dimension 
being to rescale the environment dimension $\chi\rightarrow\chi_D=\chi/D^2$. 
We note nevertheless that, although our data are consistent with (\ref{EQ:length}), one cannot rule out that some of the 
correlation lengths may saturate to a finite, although large, value. 

Related $J-Q$ models can be investigated with QMC~\cite{Lou2009} 
and $\eta_{\rm s}\simeq 0.35(2)$ and $\eta_{\rm d}\simeq 0.20(2)$ have been obtained (for the $J-Q_2$ model), which seem to deviate substantially from our estimates above. However, our estimation of $\eta_{\rm s}$ seems consistent with the VMC result~\cite{Morita2015}
$\eta_{\rm s}\sim 0.5$ obtained for the $J_1-J_2$ Heisenberg model at $J_2=0.5$. 

Note that the power-law exponent $1+\eta_{\rm d}$, extracted from the correlations at intermediate distances $d<\xi_D(\chi)$, seems to increase significantly with $D$. The predicted large value of the $D\rightarrow\infty$ dimer anomalous dimension 
might indicate that dimer correlations at the QCP or within the QCPh
are significantly suppressed compared to $J-Q$ models.

\begin{acknowledgements}
This project is supported by the TNSTRONG
ANR grant (French Research Council).  This work was granted access to the HPC resources of CALMIP supercomputing center under the allocations 2016-P1231 and 2017-P1231. DP thanks Nicolas Renon (CALMIP) and Cyril Mazauric (ATOS)  
for assistance. 
D.P. acknowledges illuminating discussions with Philippe Corboz, as well as 
helpful advices to implement the CTM algorithm.
M.M. and D.P. thank Roman Orus for insightful comments.
D.P. also acknowledges inspiring conversations with Fabien Alet, Federico Becca, Ignacio Cirac, 
Shenghan Jiang, Naoki Kawashima, Fr\'ed\'eric Mila, David Perez-Garcia, 
Frank Pollmann, Pierre Pujol,
Ying Ran, Anders Sandvik, Norbert Schuch, Frank Verstraete and Ling Wang. 
\end{acknowledgements}

\bibliography{bibliography}

\clearpage
\appendix

\section{Scaling of the $D=7$ variational energy vs inverse environment dimension}
\label{app1}

\begin{figure}[htbp]
\begin{center}
\includegraphics[width=9cm,angle=0]{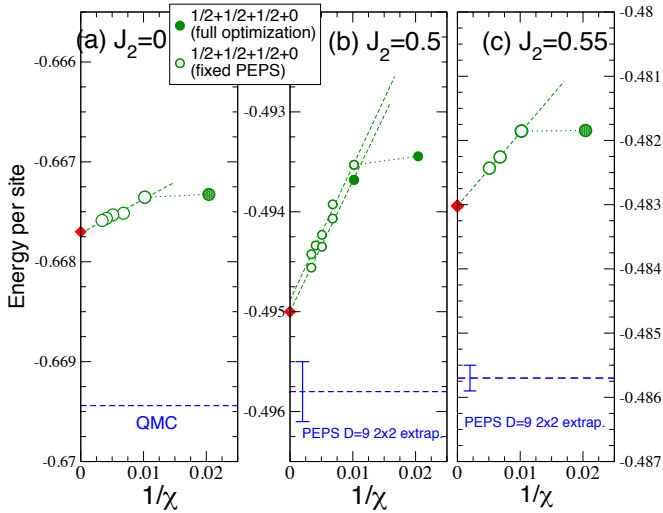}
\caption{[Color online] (a) $D=7$ iPEPS variational energies of the $J_1-J_2$ model
at $J_2=0$ (a), $J_2=0.5$ (b) and $J_2=0.55$, versus the
inverse of the environment dimension $\chi$. Full (open) symbols correspond to fully optimized (fixed) tensor
ans\"atze (see text). $\chi\rightarrow\infty$ linear extrapolations are performed using only the last data points.
Comparisons with QMC~\cite{Sandvik2010a,Sandvik2010b} and {\it finite size} $D=9$ PEPS 
extrapolations (with error bars)~\cite{Wang2016} are shown.}
\label{FIG:energy2}
\end{center}
\end{figure}

We report in Fig.~\ref{FIG:energy2}(a-c) the variational energies of the $V=\frac{1}{2}\oplus\frac{1}{2}\oplus\frac{1}{2}\oplus 0$ 
PEPS ansatz for the $J_1-J_2$ model at $J_2=0$ (unfrustrated case), $J_2=0.5$ and $J_2=0.55$. The parameters 
of the PEPS are optimized with an environment dimension $\chi_{\rm opt}=D^2=49$, independently 
for each value of $J_2$. For $J_2=0.5$ we also carried out the optimization with $\chi_{\rm opt}=2D^2=98$, providing a slightly better energy.
The environment dimension $\chi>\chi_{\rm opt}$ is then increased, keeping
the PEPS tensor fixed, and the energy is extrapolated linearly with $1/\chi$.  
At $J_2=0.5$, an excellent agreement is found with extrapolation from $D=9$ PEPS cluster update~\cite{Wang2016}. For $J_2=0$ and $J_2=0.55$ a less good agreement is found with QMC~\cite{Sandvik2010a,Sandvik2010b} and $D=9$ PEPS cluster update~\cite{Wang2016}, respectively (see text for explanation).

\section{Scaling of the $D=7$ staggered magnetization vs inverse environment dimension}
\label{app2}

\begin{figure}[htbp]
\begin{center}
\includegraphics[width=9cm,angle=0]{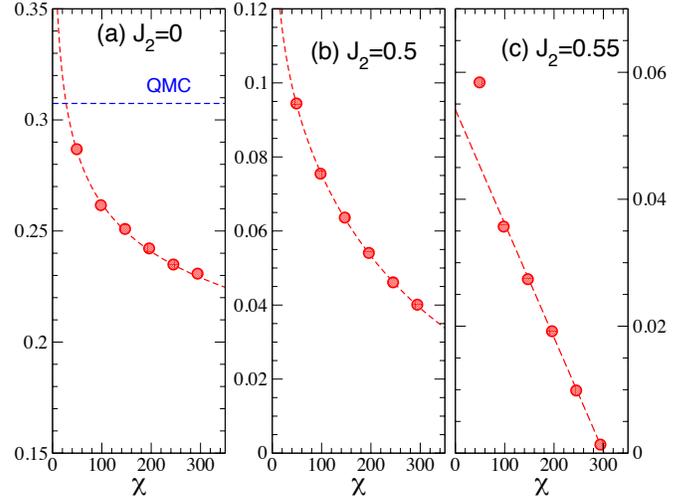}
\caption{[Color online] (a) $D=7$ iPEPS staggered magnetization of the $J_1-J_2$ model
at $J_2=0$ (a), $J_2=0.5$ (b) and $J_2=0.55$, versus environment dimension $\chi$. $\chi\rightarrow\infty$ extrapolations are 
based on power-law fits.
The exact (QMC) value of $m_{\rm stag}$~\cite{Sandvik2010a,Sandvik2010b} at $J_2=0$ is shown.}
\label{FIG:mstag}
\end{center}
\end{figure}

We report in Fig.~\ref{FIG:mstag}(a-c) the spurious staggered magnetization 
of the $V=\frac{1}{2}\oplus\frac{1}{2}\oplus\frac{1}{2}\oplus 0$ 
PEPS ansatz for the $J_1-J_2$ model at $J_2=0$ (unfrustrated case), $J_2=0.5$ and $J_2=0.55$ (optimized using $\chi_{\rm opt}=D^2=49$). 
The procedure is the same as in Appendix~\ref{app1} and the data are plotted vs $\chi$. 
For all $J_2$ values, the scaling (algebraic fits) is consistent with
vanishing $m_{\rm stag}$ when $\chi\rightarrow\infty$. Full SU(2) invariance is recovered in this case. 

\section{Extracting the correlation lengths $\xi_D(\chi)$ from the long distance correlations}
\label{app3}

\begin{figure}[htbp]
\begin{center}
\includegraphics[width=9cm,angle=0]{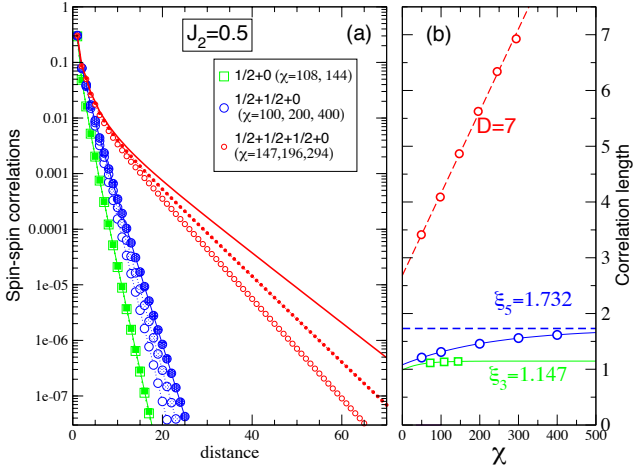}
\caption{[Color online] (a) Spin-spin correlation versus distance
for (fixed) $D=3$ , $D=5$ and $D=7$ tensors and several dimension $\chi$ of the environment (semi-log scale). 
The tensors are obtained from a full CG optimization using environment dimensions $\chi_{\rm opt}=108,100$ and $49$, respectively.
Correlation length extracted from linear fits of the asymptotic large-distance behaviors are shown
in (b) versus $\chi$. }
\label{FIG:ss}
\end{center}
\end{figure}

\begin{figure}[htbp]
\begin{center}
\includegraphics[width=9cm,angle=0]{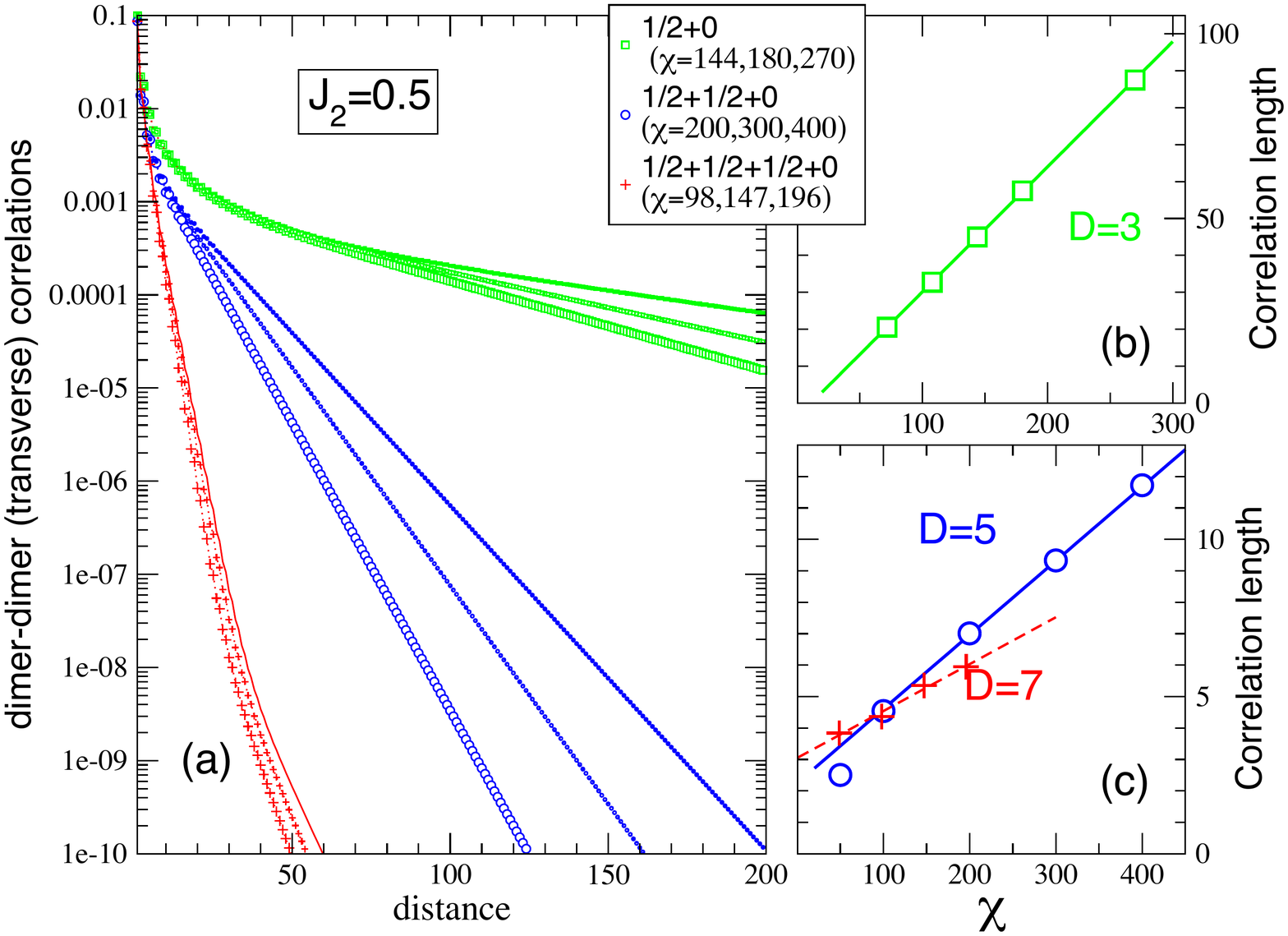}
\caption{[Color online] (a) Transverse dimer-dimer correlation versus distance
for $D=3$ , $D=5$ and $D=7$ and several values of $\chi$ (semi-log scale). 
Tensors are the same as in Fig.~\protect\ref{FIG:ss}.
Correlation lengths extracted from linear fits of the asymptotic large-distance behaviors are shown
in (b) and (c) versus $\chi$. }
\label{FIG:ddv}
\end{center}
\end{figure}

\begin{figure}[htbp]
\begin{center}
\includegraphics[width=9cm,angle=0]{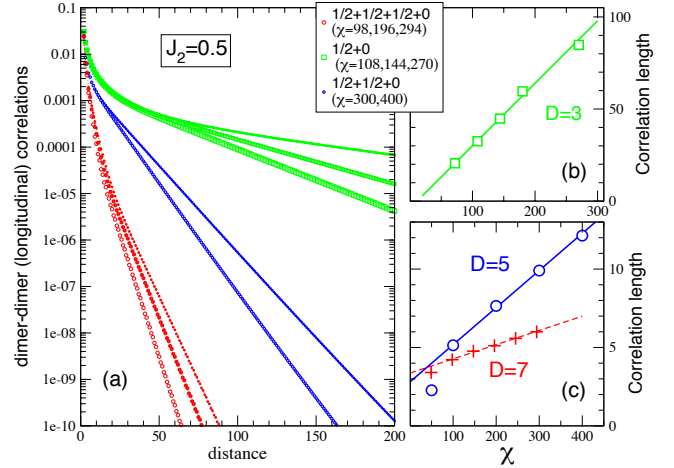}
\caption{[Color online] (a) Longitudinal dimer-dimer correlation versus distance
for $D=3$ , $D=5$ and $D=7$ and several values of $\chi$ (semi-log scale). 
Tensors are the same as in Fig.~\protect\ref{FIG:ss}.
Correlation lengths extracted from linear fits of the asymptotic large-distance behaviors are shown
in (b) and (c) versus $\chi$. }
\label{FIG:ddh}
\end{center}
\end{figure}

In order to extract the correlation lengths associated to the various correlation functions $C_\lambda (d)$ ($\lambda={\rm S, D}$) 
defined in the paper in Eqs.~(\ref{EQ:ss}), (\ref{EQ:ldd}) and (\ref{EQ:tdd}),
we have computed the long-distance correlations using the transfer matrix methods sketched in Fig.~\ref{FIG:tensor}.
Due to a finite gap in the relevant transfer matrices for all finite dimensions $D$ and $\chi$, 
one expects an exponential decay of all correlations,
$$
C_\lambda(d)\sim C_0\exp{(-d/\xi_D(\chi))}\, ,
$$ 
at sufficiently large distance $d$ (typically $d> \xi_D(\chi)$).
Let us summarize the procedure~: First, the local tensors for $D=3$, $5$ and $7$ are obtained by a full CG optimization 
(for $J_2=0.5$) using a given environment dimension $\chi_{\rm opt}=108$, 
$100$ and $49$, respectively. 
The correlations in these fixed PEPS are then computed for increasing values of the environment dimension $\chi$ in two steps~:
(i) For every choice of $\chi\ge \chi_{\rm opt}$, the new converged CTM $C$ and edge tensor $T$ are computed (by the iterative renormalization scheme) and, 
finally, (ii) used to compute the correlation functions in the setup shown in Fig.~\ref{FIG:tensor}(a-c). 
Results are displayed using 
semi-logarithmic scales in Figs.~\ref{FIG:ss}(a), \ref{FIG:ddv}(a) and \ref{FIG:ddh}(a). By fitting the asymptotic linear behaviors of the data
according to $\ln{C_\lambda(d)}=-(1/\xi) d + c_0$, one straightforwardly gets the correlation lengths $\xi$ from the slopes $-1/\xi$.

The scaling of the correlation lengths $\xi_D$ with $\chi$ are shown in 
Figs.~\ref{FIG:ss}(b), \ref{FIG:ddv}(b,c) and \ref{FIG:ddh}(b,c).
For $D=3$ and $D=5$, one observes a clear saturation of the spin correlation lengths $\xi_3$ and $\xi_5$ to rather small values 
(less than 2 lattice spacings) while the dimer correlation length scales linearly with $\chi$ suggesting that 
$\xi_D\rightarrow\infty$ in the limit $\chi\rightarrow\infty$, for which the calculation becomes exact. 
Note that the (extrapolated) spin correlation length increases with $D$ while the divergence of the dimer correlation length
becomes weaker. For $D=7$, one has to consider the {\it connected} part of the spin-spin correlation, subtracting off
the contribution from the spurious staggered spin density background. The spin correlation length no longer saturates but rather increases linearly with the 
environment dimension $\chi$. This strongly suggests that $\xi_7$ diverges in the limit $\chi\rightarrow\infty$, that
is consistent with a power-law decay of the correlation function. We believe our numerical results also support the divergence of both dimer-dimer correlation lengths, as well. Note however that,
although the transverse and longitudinal dimer-dimer
correlation lengths seem to match for $D=3$ and $D=5$, they deviate substantially for $D=7$,
which may be related to the non-vanishing of the spin-spin correlation in that case.

\section{Comparison between correlation functions in the $D=7$ PEPS}
\label{app4}

In principle, correlation lengths can also be extracted directly from the
low-energy eigenvalues of the zero dimensional transfer matrix of the one-dimensional tensor network
structures arising in Fig.~\ref{FIG:correlators}. It would be the same transfer matrix
for spin-spin and (longitudinal)  dimer-dimer correlation function, but the
difference would be how the corresponding virtual eigenvectors of
these eigenvalues transform under the symmetry. In a perfectly SU(2)-symmetric state giving rise to
a SU(2)-symmetric environment (as it occurs for $D=3$ and $D=5$), different selection rules 
for the singlet (dimer) and the triplet (spin) 
channels lead to separate blocks of the transfer matrix and, hence, to different correlation lengths, in agreement with our findings. However, for $D=7$ spontaneous SU(2) symmetry breaking occurs
and the environment acquires some (staggered) magnetization $\bf m_{\rm stag}$. 
We believe spin-rotational invariance ($U(1)$ symmetry) is still preserved around the direction of the staggered magnetization. The latter can be pointing in any (arbitrary) direction in the $(x,z)$ 
plane making difficult the symmetry analysis of the zero dimensional transfer matrix arising in Fig.~\ref{FIG:correlators}. Analysis of the correlation functions given e.g. by Eqs.~\ref{EQ:ss}, \ref{EQ:ldd} or \ref{EQ:tdd} is more straightforward. 
 
 At this point, it is not clear whether the long distance spin correlation 
described in the text is an artifact of the symmetry breaking 
that i) may lead to a mixture of (diverging) singlet and (short-range) triplet correlations or ii) may lead to "Goldstone critical behavior" of the transverse spin correlation function. 
We give arguments below that none of the above applies and argue that the critical behavior of the spin correlation function is an
intrinsic feature of the $D=7$ PEPS spin liquid. 

\begin{figure}[htbp]
\begin{center}
\includegraphics[width=9cm,angle=0]{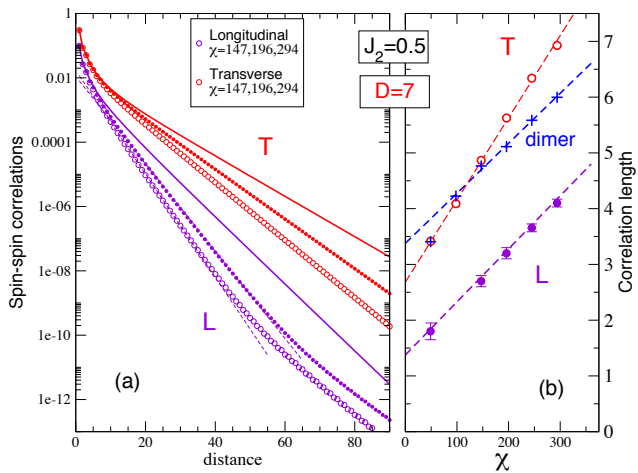}
\caption{[Color online] (a) Longitudinal and transverse spin correlations 
versus distance in the SU(2)-symmetry broken $D=7$ PEPS, for several values 
of $\chi$ (semi-log scale). 
Tensors are the same as in Fig.~\protect\ref{FIG:ss}. 
(b) Correlation lengths extracted from linear fits of the large-distance behaviors are shown versus $\chi$ and compared to the (longitudinal) dimer correlations. 
Note that, eventually, beyond some large cross-over length scale (which increases with $\chi$), the decay of the longitudinal correlation function is governed by 
the asymptotic (larger) correlation length of the transverse correlation function. 
}
\label{FIG:ssTL}
\end{center}
\end{figure}

For this purpose, we decompose the local spin operator into its longitudinal and transverse spin components,
\begin{equation}
{\bf S}_{\bf i}= S_{\bf i}^\parallel\, {\bf n}  +   {\bf S}_{\bf i}^\perp\, ,
\label{EQ:TL}
\end{equation}
where $\bf n$ is a unit vector along ${\bf m}_{\rm stag}$, $S_{\bf i}^\parallel ={\bf S}_{\bf i}\cdot {\bf n}$
and ${\bf S}_{\bf i}^\perp={\bf S}_{\bf i} - ({\bf S}_{\bf i}\cdot {\bf n})\, {\bf n}$.
The spin correlation function can be then split into its longitudinal and transverse components
as $C_{\rm s}(d)=C_{\rm s}^\parallel (d)+C_{\rm s}^\perp (d)$ with, 
\begin{eqnarray}
C_{\rm s}^\parallel(d)&=&\big< {\bf S_i}^\parallel \cdot{\bf S}_{{\bf i}+d {\bf e}_x}^\parallel\big>_0\, ,\\
C_{\rm s}^\perp(d)&=&\big< {\bf S_i}^\perp \cdot{\bf S}_{{\bf i}+d {\bf e}_x}^\perp\big>_0\, .
\label{EQ:ssTL}
\end{eqnarray}
For a true singlet wave function (for which ${\bf m}_{\rm stag}={\bf 0}$), 
whatever the choice of the vector $\bf n$, on gets $C_{\rm s}^\perp(d)=2C_{\rm s}^\parallel(d)$. 
As shown in Fig.~\ref{FIG:ssTL}(a) this is also true for the $D=7$ PEPS, at {\it short
distance} only (in semi-log scale the two curves are just shifted by $\ln{2}$). At longer distance, however, the longitudinal and transverse spin correlations show different exponential decays. As shown in Fig.~\ref{FIG:ssTL}(b) 
the correlation length of the longitudinal correlations is much shorter than the one of the transverse correlations. 
However, both seem to diverge with increasing $\chi$, suggesting that both correlators are critical, possibly power-law, in the $\chi\rightarrow\infty$ limit. This is different from a "Golstone mechanism" for which the longitudinal correlations remain short-range. Finally, we compare the two spin correlation lengths to the 
(longitudinal) dimer correlation length. Fig.~\ref{FIG:ssTL}(b) shows that none of the three (diverging) correlation length match, suggesting that 
the (supposedly) critical behavior of the spin-spin correlation is not induced by the critical behavior of the dimer correlation and is an intrinsic property of the $D=7$ PEPS.

\end{document}